\documentclass[sigconf]{acmart}
\usepackage{array} 
\usepackage{multirow}
\usepackage{graphicx}
\AtBeginDocument{%
  \providecommand\BibTeX{{%
    \normalfont B\kern-0.5em{\scshape i\kern-0.25em b}\kern-0.8em\TeX}}}

\copyrightyear{2026}
\acmYear{2026}
\setcopyright{cc}
\setcctype{by}
\acmConference[UKICER 2026]{UK and Ireland Computing Education Research 2026}{September 03--04, 2026}{Cambridge, United Kingdom}
\acmBooktitle{UK and Ireland Computing Education Research 2026 (UKICER 2026), September 03--04, 2026, Cambridge, United Kingdom}
\acmDOI{10.1145/3830800.3830803}
\acmISBN{979-8-4007-2593-7/2026/09}

\begin{document}

\title[Measuring Computational Thinking Self-Efficacy (CT-SEI)]{Measuring Computational Thinking Self-Efficacy (CT-SEI): Instrument development and preliminary evaluation}

\author[1]{Imke de Jong}
\email[1]{i.dejong1@uu.nl}
\orcid{0000-0003-0404-4011}
\affiliation{%
  \institution{Utrecht University}
  \country{The Netherlands}
}
\author[2]{Johan Jeuring}
\email{j.t.jeuring@uu.nl}
\orcid{0000-0001-5645-7681}
\affiliation{%
  \institution{Utrecht University}
  \country{The Netherlands}
}
\author[3]{Judith Masthoff}
\email{j.f.m.masthoff@uu.nl}
\orcid{0000-0001-8099-0515}
\affiliation{%
  \institution{Utrecht University}
  \country{The Netherlands}
}


\renewcommand{\shortauthors}{De Jong, Jeuring \& Masthoff}

\begin{abstract}
Much effort is put into helping students at different educational levels develop Computational Thinking (CT) skills. Self-efficacy is important for skill development. It can predict perseverance, engagement and success on educational tasks. We created an instrument to measure self-efficacy of students in higher education for the CT skills abstraction, algorithmic thinking, decomposition, evaluation and generalization. First, 91 candidate items were created by including, adapting and extending items found in the literature. These items were evaluated by experts in the field of CT and education. 54 items remained and to reduce the number of items further, data was collected from 270 students in higher education recruited both through Prolific and a university setting in Costa Rica. Through principle component analysis (PCA) using a subset of 200 responses, the number of items was reduced to 27. Confirmatory factor analysis (CFA) using the remaining responses in the dataset showed the items can be divided into two categories: (1) creating the solution and (2) evaluating the solution. The created instrument can be valuable when assessing CT self-efficacy of students in higher education. With additional validation (e.g. examination of test-retest validity), we believe the scale could be used to evaluate the effectiveness of interventions, or decide what interventions should be provided to foster CT skill development. 
\end{abstract}



\maketitle

\section{Introduction}
Computational Thinking (CT) skills are widely recognized as important 21st century skills that may be beneficial to everyone \cite{RefWorks:doc:5d690b65e4b08d1f63d98acd}. Although different typologies exist, CT often refers to skills like algorithmic thinking, abstraction and decomposition. These skills help students and practitioners in different domains to design computations and leverage computational tools \cite{denning2019computational}. With the rise of generative AI tools such as Github Copilot and ChatGPT, good command of CT skills may become even more important: better CT skills correlate with better skills in employing these tools \cite{jeuring2023skills}. 

On different educational levels, from K-12 to higher education, teachers design interventions for CT skill development  \cite{de2020computational,RefWorks:doc:5d67e6d7e4b08d1f63d961c4}. The effectiveness of these interventions can be depend on a student's self-efficacy; a psychological construct that influences skill development and task performance. Self-efficacy is a predictor for perseverance, engagement and success on educational tasks \cite{schunk2012learning}. Bandura defines it as "\emph{a belief about what one can do under different sets of conditions with whatever skills one possesses}" and as "\emph{people’s beliefs in their capabilities to produce given attainments}" \cite{bandura_1997}. Self-efficacy does not refer to someone's belief in possessing a particular skill (e.g., I possess the skill algorithmic thinking), but it refers to whether they believe they can complete a given task or use a skill in a particular context (e.g., I can create an algorithm to solve a programming task) \cite{bandura_1997}. A student's CT self-efficacy could inform what CT interventions and instruction should be employed \cite{TANG2020103798}. The benefits of tailoring interventions to self-efficacy 
have been demonstrated in CS contexts. For example, \citeauthor{10.1145/3631802.3631832} \cite{10.1145/3631802.3631832} showed that providing Parsons problems when learning 
programming concepts is more effective for students with low self-efficacy.

In this study we focus on the development of CT skills in higher education and create an instrument to measure students' self-efficacy for different CT facets. We aim to develop an instrument that can be applied in different contexts. However, because context is important when discussing self-efficacy, we focus on examining CT skills in the context of a Python programming task. We selected the Python programming language because it is often used in introductory programming courses \cite{10.1007/978-3-030-71782-7_31}. Our study is guided by the following research question: \textit{How can Computational Thinking self-efficacy be measured in the context of a Python programming task?} 

\section{Related work}
Measuring CT self-efficacy is not trivial. Self-efficacy strongly depends on the specific situation or context in which one needs to perform a task or use a skill \cite{bandura_1997}. To effectively measure self-efficacy, the domain and the capabilities necessary to complete a task successfully have to be well defined \cite{bandura_2006}. This brings challenges when creating an instrument to measure CT self-efficacy. Although being able to think computationally is deemed important, what this entails exactly is still being debated. Different definitions for CT have been proposed through the years \cite{RefWorks:doc:5d55333ae4b0b937c802dd9f,RefWorks:doc:5d67e6d7e4b08d1f63d961c4,RefWorks:doc:5d553701e4b0b937c802de02}. 
The competences, abilities or skills that are part of CT (and thus the capabilities that make someone a good computational thinker) are also subject to debate. Based on different studies, \citeauthor{RefWorks:doc:5d68e305e4b02c439d835fd7} \cite{RefWorks:doc:5d68e305e4b02c439d835fd7} identify 19 different competences involved in CT, among which abstraction, data analysis, pattern recognition, algorithm design, debugging and error detection, and problem solving. \citeauthor{RefWorks:doc:5d55333ae4b0b937c802dd9f} \cite{RefWorks:doc:5d55333ae4b0b937c802dd9f} combine the results of different studies and conclude that CT consists of competencies in abstraction, decomposition, algorithms, evaluation and generalization. Meanwhile, \citeauthor{RefWorks:doc:5d67e6d7e4b08d1f63d961c4} \cite{RefWorks:doc:5d67e6d7e4b08d1f63d961c4} identify six general facets: decomposition, abstraction, algorithms, debugging, iteration and generalization. The Computer Science Teachers Association and the International Society for Technology in Education propose CT also encompasses dispositions and predispositions like confidently dealing with complexity, handling open-ended questions, and perseverance when solving difficult problems \cite{RefWorks:doc:5d8e0c73e4b0f7862331a77b}. In this study we follow the refinement of \citeauthor{dagiene2017developing} \cite{dagiene2017developing} of \citeauthor{selby2015relationships}'s proposal \cite{selby2015relationships}, see Table \ref{tab:CTdefinitions}. The five skills identified by \citeauthor{selby2015relationships} \cite{selby2015relationships} (abstraction, algorithmic thinking, decomposition, evaluation, and generalization) are often mentioned when CT is discussed. \citeauthor{dagiene2017developing}'s refinement \cite{dagiene2017developing} offers a level of concreteness to the skills that is beneficial when examining self-efficacy, effectively specifying three sub-skills for each main skill. 
We view CT as a problem-solving strategy that encompasses skills that are important in the context of computer science but are also applicable in other domains \cite{RefWorks:doc:5d553701e4b0b937c802de02}.

\begin{table}[]
    \caption{CT skill definitions}
    \label{tab:CTdefinitions}

    \begin{tabular}{p{4.0cm}p{4.0cm}}
    \toprule
        \textbf{Skill} \cite{selby2015relationships} & \textbf{How to identify}  \cite{dagiene2017developing}\\
    \midrule
    \textbf{Abstraction}: The ability to decide what details of a problem are important and what details can be ignored & Removing unnecessary details (RUD); Spotting key elements in problem (SK); Choosing a representation of a system (CAR) \\ \midrule
    \textbf{Algorithmic thinking}: Devising explicit instructions for accomplishing tasks and creating step-by-step set of instructions& Thinking in terms of sequences and rules (TIS); Executing an algorithm (EA); Creating an algorithm (CA)\\ \midrule
    \textbf{Decomposition}: The ability to break problems down into smaller, more easily solved, parts & Breaking down tasks (BDT); Thinking about problems in terms of component parts (CP); Making decisions about dividing into sub-tasks with integration in mind, e.g. deduction (INT)\\ \midrule
    \textbf{Evaluation}: The ability to evaluate processes, in terms of efficiency and resource utilisation, and the ability to recognise and evaluate outcomes & Finding best solution (FBS); Making decisions about good use of resources (MDR); Determining fitness for purpose (DFP)\\ \midrule
    \textbf{Generalization}: The ability to express a problem solution in generic terms, which can be applied to different problems that share some of the same characteristics as the original & Identifying patterns as well as similarities and connections (IPS); Solving new problems based on already-solved problems (SNP); Using the general solution, e.g. induction (UGS)\\
    \bottomrule
    \end{tabular}
\end{table}

We are not the first to create an instrument to measure CT self-efficacy. 
\citeauthor{boulden2021measuring} \cite{boulden2021measuring} study self-efficacy in the context of teaching CT principles, and created the T-STEM CT-scale. This scale can be used to measure self-efficacy in relation to teaching CT skills and practices of in-service teachers, but does not measure self-efficacy for the CT skills themselves. 
Other scales contain questions that target CT skills more directly, such as the studies from \citeauthor{kukul2019computational} \cite{kukul2019computational}, \citeauthor{gulbahar2019self} \cite{gulbahar2019self}, and \citeauthor{KORKMAZ2017558} \cite{KORKMAZ2017558}. We believe these scales need to be refined and extended to effectively measure CT self-efficacy. The items in these scales are mostly generically phrased, and often refer to the ability to solve "a problem", without specifying a context. To effectively measure self-efficacy, however, a context is important \cite{bandura_1997}, e.g., describing the problem one has to solve using the skills. 
Furthermore, some of the items in the existing scales may not be appropriately phrased to measure self-efficacy. These questions target knowledge or ask whether one \textit{would} take a particular action, whereas self-efficacy concerns if someone believes they \textit{could} take action. If someone would  perform a particular action is not relevant for self-efficacy, neither is their estimate of the knowledge they have \cite{bandura_2006}. Statements should thus be phrased as “I can do” not “I will do”. Our aim is to combine, refine and extend existing self-efficacy scales with these aspects in mind, creating a comprehensive CT self-efficacy instrument (CT-SEI). 

\section{Development of the CT Self-Efficacy instrument}
Two principles for measuring self-efficacy \cite{bandura_1997, bandura_2006} guide the setup of the instrument we propose. First, a difficulty level, situation or domain should be specified when measuring self-efficacy, as someone's self-efficacy depends on the context of the task. Self-efficacy might be high in one setting but low in another. Second, when measuring self-efficacy for skills, you should also inquire whether the problem solver believes they can successfully complete the task in which the skills should be used. According to \citeauthor{bandura_1997} \cite{bandura_1997} this is especially important in situations where there is still debate about the specific behaviors or skills that lead to a successful outcome on a task. This is the case for situations in which CT skills are involved, as evidenced by the different definitions that exist for CT. 
Because of these two principles, the proposed instrument consists of three parts:
\begin{enumerate}
    \item An introduction that sets the context and describes a task. This part contains an example assignment, or a description of a type of assignment or problem. We used the following (verbatim) Python programming assignment as context: \textit{"Imagine you are taking a programming course. For one of the assignments, you are asked to program a guessing game. The game has a broad target audience. Children could for example use it to decide who will be "it" in a game of tag, adults could use it to decide who will unload the dishwasher, etc. In the game, the player guesses a randomly generated number between 1 and 10 times the number of players. To clarify: with 2 players, the generated number should be between 1 and 20; with 3 players, the generated number should be between 1 and 30, etc. The first player to guess the correct number wins the game. There is a variable number of players, and each player is allowed to guess 6 times (although not consecutively, of course). After each guess, the program will indicate whether the guessed number is correct or incorrect. If the guess is incorrect, it will indicate whether it is too high or too low, and ask the next player for their guess. If the number wasn't guessed after every player guessed 6 times, the program should output the correct answer and randomly pick a winner. For the current assignment, you are not required to program a GUI. The input can be given through the console."}.
    \item A question about the self-efficacy for the task or assignment: \textit{”How confident are you in your ability to solve such a puzzle / such a problem / such an assignment?”}. This question can be adjusted to the context. Participants indicate their self-efficacy on a scale from 0 to 10, with the following descriptions: 0 = cannot do at all, 5 = moderately can do, 10 = highly certain can do. This part was added to measure whether a student feels they are proficient in the task itself, regardless of feeling able to use CT skills for the task.
    \item A set of self-efficacy statements for each CT skill, preceded by \textit{”When I solve such a puzzle / such an assignment / such a problem...”.} This statement can be adjusted to the context. Participants indicate their self-efficacy on a scale from 0 to 10, with the following descriptions: 0 = cannot do at all, 5 = moderately can do, 10 = highly certain can do. These statements cover all aspects of CT as presented in Table \ref{tab:CTdefinitions}.
\end{enumerate}

To create the third part of the instrument, the CT self-efficacy statements, we employed two steps: item development and item selection. We discuss these steps in the next sections.

\subsection{Step 1: Item development}
We created a set of statements or items that could be used to measure self-efficacy. They cover the five CT skills and their sub-skills as presented in Table~\ref{tab:CTdefinitions}. Existing CT self-efficacy assessments \cite{kukul2019computational, gulbahar2019self, KORKMAZ2017558} served as inspiration and initial starting point. We categorized items from these scales under the appropriate CT skills and reformulated them to an "I can" format. We rephrased the items so they could be adapted to different contexts. Because the items from the existing scales did not cover the definitions of all skills, we formulated additional items. Finally, to increase the item pool, we created alternatives with slightly different wording for each of the items created up to this point. This resulted in a set of 91 items (18 items for Abstraction, 18 for Algorithmic Thinking, 16 for Decomposition, 20 for Evaluation, and 19 for Generalization)\footnote{The full set of items is available from the first author upon request}.

We examined the content validity of the items by asking independent experts (not part of the author team) for their opinion on the appropriateness of the items for measuring CT self-efficacy. We consulted experts in both Educational Sciences and CT for this validation. One educational researcher evaluated the proposed set-up of the instrument and the wording of the items. The expert judged the items and the setup appropriate for measuring self-efficacy. Three researchers in the field of CT individually evaluated the proposed items on their connection with the topic of CT. In a self-paced evaluation, the experts first read the definitions of the skills used in the study as presented in Table \ref{tab:CTdefinitions}. We then presented all candidate items in random order, and the experts judged whether or not items should be included. We selected the items that all three experts deemed appropriate for inclusion. This reduced the pool of items to 54 (10 items for Abstraction, 12 items for Algorithmic thinking, 10 items for Decomposition, 9 items for Evaluation, and 13 items for Generalization). In this set, each skill and sub-skill was represented by multiple items. 

\subsection{Step 2: Item selection} 
\label{scale_development}
We collected data to remove highly similar items, shorten the time needed to administer the instrument and examine if the number of items could be reduced without decreasing the reliability of the item set. We applied principle component analysis (PCA) to analyze the data. Through confirmatory factor analysis (CFA) we inspected the relationship between the resulting item set. 

\subsubsection{Data collection}
To collect data from higher education students, we recruited participants through the online platform Prolific and supplemented this with 21 responses gathered from students that took a Python programming course at an institute for higher education in Costa Rica. The questionnaire was translated to Spanish for the latter. We collected data in 2023 and 2024. We specified two inclusion criteria for participant selection for Prolific: the participant should (1) be currently enrolled as student, and (2) have experience with programming in Python. Demographic data (gender, age and nationality) was received through Prolific, along with details about which programming languages (out of a list of 18) the participants feel proficient in. 249 participants responded to the questionnaire distributed on Prolific, with 41 different nationalities. The largest portion of respondents was from South Africa (37 percent).
For the item reduction process through PCA, we used random case selection in SPSS 28 to select 200 (121 male, 79 female) out of 249 Prolific responses. The remaining 49 responses were combined with the 21 additionally collected responses to examine the factor structure of the item set through CFA (see Section \ref{sect:cfa}).
Table \ref{tab:descriptivesPCAandCFA} gives descriptives of the participants (age, number of programming languages they are proficient in, and self-efficacy for the example assignment - part 2 of the instrument). Details of three participants are omitted due to non-disclosure of age or gender.

\begin{table*}[]
 \caption{Descriptives of participants included in PCA (\textit{n male = 121, n female = 79)} and CFA (\textit{n male = 44, n female = 23)}}
 \label{tab:descriptivesPCAandCFA}
\begin{tabular}{lllllllll}
\toprule
              & \multicolumn{4}{l}{PCA dataset} & \multicolumn{4}{l}{CFA dataset} \\
              & Min  & Max & Mean & SD & Min  & Max  & Median  & SD \\
\midrule
Age           &  19    &   60  & 26.2      &  6.96  &   18   &  54     & 24.5        &  6.46   \\
Programming languages   &  1    & 18     &   4.86     &  3.68   &  1    & 18     & 4.4        &  3.39   \\

Assignment SE  &  1    &   10   &  7.3      &  2.37  &  2    & 10      &    8.1     & 2.06   \\
\bottomrule
\end{tabular}%
\end{table*}

\subsubsection{Reducing the number of items} PCA with oblimin rotation (Kaiser maximization) was used to reduce the item set. PCA provides a mechanism to see how many factors a set of items measures, and how items load on these factors \cite{matsunaga2010factor}. We chose to employ PCA instead of Exploratory Factor Analysis, as our candidate items are based on a theoretical model consisting of five factors (abstraction, algorithmic thinking, decomposition, evaluation and generalization). We analysed each of these factors separately, with the goal of identifying whether the created items indeed load upon one common factor. For all five item sets, the KMO and Bartlett's test revealed the data was suitable for PCA (KMO > .900 and Bartlett's test of spherity significant with p <0.01). The oblique-rotation method oblimin was chosen, because we expected that the items would be interrelated, and Spearman's correlation analysis showed strong correlations between the items (r(200) > 0.4). 

For algorithmic thinking, decomposition, evaluation and generalization, the results of the PCA showed all items load on one factor with factor loadings of .729 or higher. 
For abstraction the PCA identified not one but two components. Although most items loaded on one factor, two items strongly loaded on a different second factor (loadings of respectively 0.915 and 0.878). Two other items also loaded on this factor, but with less strong loadings: 0.308 and 0.670. We decided to remove the items with the strongest loading on the second factor from the analysis. Examination of these items showed they were worded in a complex way. Both inquired about a participant's ability to ignore unimportant details, using double negation. This may have influenced the way participants responded to the items. This also becomes apparent when looking at the mean scores of these items, which is almost 1 point lower than the rest of the items. After removing the items from the analysis, the PCA resulted in the identification of one factor, with loadings of 0.639 or higher.

We reduced the number of items per skill by examining the factor loadings and the wording of the items, ensuring each of the sub-skills from Table \ref{tab:CTdefinitions} was represented in the final set.  Selection was based on the factor loadings and wording of the items. In principle, the alternative with the higher loading was selected. However, in nine cases the first two authors agreed after examining the items that the item with a lower factor loading was worded more clearly than the alternative item. The difference between factor loadings was less than 0.065 in these cases. The item reduction process resulted in 27 items. The final set of items can be found in Table \ref{tab:CFA_factorloadings}. The number of items per skill differs for two reasons. First, the definitions of the sub-skills sometimes contain multiple elements in one sub-skill (e.g. "thinking in terms of \emph{sequences} and \emph{rules}"). Separate items were developed for these elements. Second, experts on CT deemed these items appropriate for inclusion. We kept items measuring something that is not covered by one of the other items. We checked reliability and internal consistency of the set of items as a whole and for the five individual skills. Cronbach's $\alpha$ for the 27 items combined was .976, pointing to very high internal consistency. Cronbach's $\alpha$ for each of the skills also indicates high reliability: $\alpha$ = .872 for abstraction, $\alpha$ = .924 for algorithmic thinking, $\alpha$ = .927 for decomposition, $\alpha$ = .846 for evaluation, and $\alpha$ = .928 for generalization. 

\subsubsection{Examining the factor structure}
\label{sect:cfa}
To assess how well the items fit the theoretical model of five CT skills we used CFA with robust maximum likelihood, using the R package lavaan (estimator = MLM, std.lv=TRUE). Based on the theoretical definition of CT and the PCA analysis, we examined a number of different models. The analysis was done using the data of the remaining 70 respondents. Because the data is not normally distributed we chose the robust maximum likelihood method \cite{brown2006confirmatory}. CFA uses different fit statistics to evaluate the fit of the data to the model. We report the model chi-square ($\chi^2$), absolute fit index (RMSEA) and comparative fit indices (CFI and TLI). The indices should be smaller than 0.06 (RMSEA), larger than 0.95 (CFI and TLI), and $\chi^2$ should be insignificant \cite{brown2006confirmatory}. 

First, we examined the theoretical model (Model 1) as presented in Table \ref{tab:CTdefinitions}. The statements in the final item set incorporate all skills and sub-skills presented there. However, a CFA analysis with this factor structure did not lead to a valid model. The results showed a strong relationship between algorithmic thinking and decomposition, and algorithmic thinking and evaluation; both resulting in a computed covariance higher than 1. To clarify the factor structure further, we examined the theoretical relationship between these skills and decided to combine the items for algorithmic thinking and decomposition into one factor. This four-factor model (Model 2) was a valid model (see Table \ref{tab:CFA_results}). Finally, we examined a third model (Model 3) based on an inspection of the items included in the final item set. In this model we separate the items into two factors. The first factor relates to creating the solution to a problem, and consists of 3 sub-factors: (1) abstraction, (2) generalization, and (3) a combination of algorithmic thinking and abstraction. The second factor relates to evaluating the created solution, and incorporates the items created for the skill evaluation. This third model shows a slightly improved fit. Table \ref{tab:CFA_factorloadings} shows this factor structure and the loadings of the items on their factors.

\begin{table}[t]
        \caption{CFA results. Model 1 was not computable}
    \begin{tabular}{p{2cm}cccc}
         \toprule
         \textbf{Model} & \textbf{$\chi^2$} & \textbf{RMSEA} & \textbf{CFI} & \textbf{TLI} \\
         \midrule
         Model 1  & n/a & n/a  & n/a & n/a \\
         Model 2  & 349.393, p = 0.109 & 0.054 & 0.967 & 0.964 \\
         Model 3 & 346.754, p = 0.146 & 0.050 & 0.972 & 0.969 \\
         \bottomrule
    \end{tabular}
    \label{tab:CFA_results}
\end{table}

\begin{table*}
\caption{Final selection of statements, including CFA factor loadings for model 3.}
\begin{tabular}{lp{9cm}c}

\toprule
            ID*          & Statement  & Factor loading \\

\midrule
\multicolumn{3}{l}{\textbf{Factor 'Create solution'}}\\
\multicolumn{3}{l}{\textit{Abstraction}}\\
AB\_1\_SK&I can select the information I need for the solution while reading the problem description&0.814\\
AB\_2\_SK&I can distinguish between information necessary and unnecessary for the solution while reading the problem description&0.808\\
AB\_3\_SK&I can focus on the important details of the problem&0.854\\
AB\_4\_RUD&I can pay attention to the important details of the problem while ignoring unimportant details&0.800\\
AB\_5\_CAR&I can represent the problem in a way that helps me solve it (e.g. by drawing a figure, making a table, etc.)&0.787\\
&&\\
\multicolumn{3}{l}{\textit{Algorithmic Thinking \& Decomposition}}\\
AT\_1\_TIS&I can think of the solution in terms of a sequence of steps&0.880\\
AT\_2\_TIS&I can think of the solution as a set of rules to follow&0.771\\
AT\_3\_EA&I can follow a step-by-step procedure&0.843\\
AT\_4\_CA&I can create simple step-by-step procedures&0.830\\
AT\_5\_CA&I can create conditional step-by-step procedures (if.. then.. else...)&0.856\\
AT\_6\_CA&I can create step-by-step procedures with loops (repeating recurring patterns)&0.810\\
DC\_1\_BDT&I can break the problem down into smaller parts to solve it&0.874\\
DC\_2\_BDT&I can break the problem down into sub-problems that are easier to solve&0.809\\
DC\_3\_CP&I can understand whether the problem is composed of sub-problems&0.858\\
DC\_4\_INT&I can break the problem down into smaller parts, while keeping integration between the parts in mind&0.814\\
DC\_5\_INT&I can understand how different parts of the problem are related&0.874\\
&&\\
\multicolumn{3}{l}{\textit{Generalization}}\\
GE\_1\_SNP&I can think about problems I have solved before and relate them to the current problem according to their similarities and differences&0.877\\
GE\_2\_SNP&I can think about problems I have solved before and use similarities and differences to solve the current problem&0.888\\
GE\_3\_SNP&I can understand how a problem I have encountered differs from the problems I have encountered before&0.905\\
GE\_4\_IPS&I can look for recurring, generic patterns in the problem&0.802\\
GE\_5\_IPS&I can evaluate whether a problem I have solved before is similar&0.850\\
GE\_6\_UGS&I can generalize the solution I created for a problem to other problems&0.693\\
GE\_7\_UGS&I can use generic solutions&0.743\\
&&\\
\midrule
\multicolumn{3}{l}{\textbf{Factor 'Evaluate solution'}}\\
\multicolumn{3}{l}{\textit{Evaluation}}\\
EV\_1\_FBS&I can try to find out if there is another solution after having already solved the problem&0.756\\
EV\_2\_FBS&I can try to find alternative solutions that are actually a better solution to the problem&0.611\\
EV\_3\_MDR&I can evaluate whether I have found the most efficient solution&0.561\\
EV\_4\_DFP&I can decide whether the solution I chose meets the objectives&0.921\\
&&\\
\bottomrule
\multicolumn{3}{l}{*the abbreviations used for the last part of the ID refer to the sub-skills as presented in Table \ref{tab:CTdefinitions}}\\
\label{tab:CFA_factorloadings}
\end{tabular}
\end{table*}

\subsection{CT Self-efficacy gender differences}
Previous studies have shown differences in self-efficacy between women and men in subjects related to computer science \cite{huang2013gender}. We examined our data for gender differences through Mann-Whitney U tests, as the data was not normally distributed and the sample size was small. The dataset used for the PCA contained no significant differences for age (U = 4735.50, p = .912) or number of programming languages (U = 4506.50, p = .492). However, for the overall self-efficacy regarding the assignment (Part 2), men scored significantly higher than women (U = 3853.00, p <.05), with a mean of 7.57 for men and 6.87 for women.
We also examined the differences between male and female participants in their answers to each of the self-efficacy items. For the 54 items in the PCA, male participants displayed higher self-efficacy than female participants for 53 items, with differences between 0.11 and 1.11 on a scale of 0 to 10. Of these differences, 30 were significant (p<.05). Female participants indicated higher self-efficacy for the item "\emph{I can evaluate whether I have found the most efficient solution}" (EV\_3\_MDR) on evaluation, with a difference of 0.04. This difference was not significant.

We repeated this analysis with the data and items of the CFA. Again, no significant differences were found for age (U = 403.00, p = .172) or number of programming languages (U = 470.00, p = .447). Although overall self-efficacy was again lower for female (M = 7.61) than male (M = 8.32) participants, this difference was not significant (U = 388.50, p=.064). Male participants had a higher self-efficacy score than female participants for 25 out of the 27 CT self-efficacy items. For three items this difference was significant (p <.05) according to a Mann-Whitney U test (AT\_5\_CA, DC\_5\_INT, and EV\_4\_DFP). For two evaluation items, self-efficacy of female participants was higher compared to male participants (EV\_3\_MDR and EV\_2\_FBS), but these differences were not significant. 

\subsection{Limitations}
Our research has a number of limitations. First, our instrument is context-dependent: we examined CT self-efficacy in the context of Python programming. We expect the instrument to provide valid results in other contexts as well when adjusted to that context, but this should be validated in future studies.
Second, for creating a scale, it is advised to have many participants answering the initial set of items, preferably in a ratio of 10 participants per item \cite{10.3389/fpubh.2018.00149}. We used the answers of 200 participants for the PCA, but the CFA was done with just 70 responses. Although 
the results of factor analysis can be stable with smaller sample sizes \cite{maccallum1999sample} and we 
gathered data from two different test panels, we would like to reexamine the factor structure with data from a larger number of participants. The instrument's robustness should be investigated further by collecting data from more participants, examining different populations, and by examining test-retest reliability.

\section{Discussion and Conclusion}
We have presented the development of a CT self-efficacy instrument that can be used to measure students' perceptions of their CT skills in the context of a programming assignment. We created items by examining the literature and existing scales. Experts evaluated the setup of the instrument and the items to determine their validity for measuring CT self-efficacy. Analyzing data from 270 participants, we created an instrument consisting of three parts. The first part contains the description of an assignment and provides context for the self-efficacy assessment. The second part assesses the overall self-efficacy of a participant for this assignment. The third part consists of 27 statements covering self-efficacy for the use of CT skills in the context of the assignment. We used CFA to examine the factor structure of the items. Two models show a good model fit, with one pointing towards a distinction between creating and evaluating a solution for a problem.

We found differences between the self-efficacy of men and women. 
On average, women indicated lower self-efficacy than men. This 
is in line with studies that examined self-efficacy in 
computer science~\cite{huang2013gender, 10.1145/3626252.3630811} and has implications for the representation of women in the field. 
For example, \citeauthor{10.1145/3501385.3543963} \cite{10.1145/3501385.3543963} showed  a decreased likelihood of woman enrolling in future CS courses due to lower self-efficacy. 
Differences between genders on perceived computational thinking proficiency have also been found. For example, \citeauthor{10.3389/feduc.2024.1478666} \cite{10.3389/feduc.2024.1478666} found significant differences in algorithmic thinking. Gender differences can have important implications for interventions that aim to develop computational thinking skills of all students and for those aiming to increase the representation of women in 
computing sciences. These efforts might benefit from interventions that cover the domain of interest (e.g., Python programming) and take developing and supporting a student's CT self-efficacy into account, for example by providing additional support or explanations for students with low self-efficacy.

Table \ref{tab:CFA_factorloadings} shows that some skills or sub-skills are represented by more items than others in the final instrument. Looking at the CFA results and the high Cronbach's $\alpha$ values, the number of items per factor could have been further reduced to achieve a more balanced distribution. However, we decided to keep all items to preserve alignment with the current theoretical definition of CT. Although beyond the scope of this study, in light of discussions about the definition of CT it would be interesting to examine the close relationship between the factors.
When using the instrument to assess a student's CT self-efficacy and guide selecting interventions it could be valuable to look at the results of the identified factors and items separately instead of one combined CT self-efficacy score. Individual factors and items can show what to focus on for a particular student, while a single self-efficacy score obscures this. 

With additional validation, the final instrument can be valuable for anyone wanting to measure CT self-efficacy. For educators, a student’s CT self-efficacy could inform the selection of CT interventions and the focus of instruction. The scale can also be used as a measurement instrument by researchers interested in students' perceptions of their CT skills. The items are validated by experts, cover the CT skills in detail and are designed following the principles of measuring self-efficacy. The setup of the instrument allows for an in-depth assessment of CT skills self-efficacy in context. Although in the current study the context was a Python programming assignment, the instrument can be adjusted to fit a different type of assignment for which CT skills are important. This requires changing the first part of the instrument. For example, the context could be changed to a non-programming assignment like a Bebras task.  As next steps, we are running an intervention study to investigate what the effect of worked examples or practice problems is on students CT self-efficacy as measured with the instrument. We also plan to examine CT self-efficacy in relation to students' grades in subjects where CT skills are used. We want to investigate if low CT self-efficacy is reflected in the grades of the students following a Python programming course.

\begin{acks}
We thank the experts for evaluating the items, and all participants who reflected on their CT skills by filling out the survey. We also thank Isaac Alpizar Chacon for translating the instrument.
\end{acks}

\bibliographystyle{ACM-Reference-Format}
\bibliography{references}

\end{document}